# Current-induced nonreciprocity and refraction-free propagation in a one-dimensional graphene-based photonic crystal

D. P. Huang and K. Y. Xu

*School of Physics and Telecommunication Engineering, South China Normal University, China*

**Abstract**

Nonreciprocal photonic devices play a significant role in regulating the propagation of electromagnetic waves. Here we theoretically investigate the nonreciprocal properties of transverse magnetic modes in a one-dimensional graphene-based photonic crystal subjected to an applied electrical DC bias. We find that drifting electrons driven by the external DC electric field can give rise to extremely asymmetric dispersion diagrams. Furthermore, when the drifting electrons travel antiparallel to the normal component of the incident wave vector, the negative refraction is strongly suppressed, causing the energy of light to flow along the direction of the direct electric current. Our theoretical findings can be used to design nonreciprocal nanophotonic devices and enable light to propagate without refraction.

1. **INTRODUCTION**

Optical nonreciprocity can be achieved by breaking time-reversal symmetry (Lorentz reciprocity) in photonic systems [1-3]. Nonreciprocal photonic devices based on time-reversal symmetry breaking play a significant role in controlling light propagation. In general, such devices can maintain the light propagation in one direction but hinder it in the opposite direction [4,5]. There are various methods for time-reversal symmetry breaking, including the use of optical nonlinearity [5-9], spatiotemporal modulations [11-16] and magneto-optical effects [17-31]. Very recently, the issue of truly unidirectional surface plasmon polaritons (SPPs), based on magneto-optical effects, has attracted much attention. According to the Ref. [26], if nonlocal effects and a realistic material model were taken into consideration, truly unidirectional SPPs cannot exist in the nonreciprocal plasmonic systems. However, in a follow-up theoretical study, the authors found that, a class of one-way propagating SPPs described by an extremely asymmetric dispersion located in the bulk mode bandgap, can robustly exist at the interface between an opaque material and a magnetized plasma, even though nonlocal effects are included [29]. This shows that the nonlocality may play an important role in studying the topic of nonreciprocal responses, but the inclusion of nonlocal effects does not hinder the existence of truly unidirectional SPPs in some cases.

Recently, nonreciprocal responses in plasmonic systems have been realized by applying an external DC voltage [32-49]. Such a novel approach can lead to a Doppler frequency shift of SPPs [35,37], wherein forward-propagating (backward- propagating) waves traveling parallel (antiparallel) to the direction of drifting electrons obtain a blueshift (redshift) and the corresponding propagating length is enhanced (suppressed) [36,37]. Hence the dispersion diagram of the SPPs will become extremely asymmetric, resulting in nonreciprocal responses. However, the effect of the drifting electrons on a one-dimensional (1D) graphene-based photonic crystal has, to the best of our knowledge, not been explored.

In this work, we investigate the nonreciprocal properties of transverse magnetic (TM) modes in a 1D graphene-based photonic crystal subjected to an electrical DC bias. Using the transfer matrix method, we find that the band structure of TM modes and the corresponding hyperbolic isofrequency contours become more and more asymmetric as drift velocity increases. Moreover, we show that drifting electrons driven by the DC electrical field can suppress the refraction angles of the wave vector and group velocity. Such properties may offer a novel route to control light propagation.

## 2. THEORETICAL MODEL

Consider a periodic multilayer structure consisting of graphene sheets and dielectric materials as shown in Fig. 1. Since the transverse electric (TE) modes maintain time-reversal symmetry in the presence of an external bias and the corresponding dispersion lacks hyperbolic-like characteristic [42-44], here we mainly focus on the TM-polarized waves. By using the transfer matrix method and applying the Bloch theorem, the corresponding dispersion relation for TM-polarized waves can be written as [45,46]

$$\cos(K_B d) = \cos(k_z d) - \frac{i\sigma_g^{drift} k_z}{2\omega\varepsilon_0 \varepsilon_d} \sin(k_z d) \qquad (1)$$

where $K_B$ is the Bloch wavevector, $d$ is the period, $\varepsilon_0$ is the vacuum permittivity, $\varepsilon_d$ is the relative permittivity, $k_z = \sqrt{\varepsilon_d (\omega/c)^2 - k_x^2}$, $k_x$ (assuming $k_y = 0$) is the $x$ component of the wavevector in the dielectric medium, and $\sigma_g^{drift}$ is the nonlocal conductivity of the current-carrying graphene. In the absence of a DC bias voltage, the general nonlocal conductivity of graphene can be obtained by employing the random-phase approximation (RPA) [47-49]. Then we have

$$\sigma_g(\omega, k_x) = i\omega \frac{e^2}{k_x^2} \Pi(\omega, k_x), \qquad (2)$$

where $\Pi(\omega, k_x) = \frac{k_x^2}{4\pi\hbar\sqrt{\omega^2 - k_x^2 v_F^2}} \left[ G\left(\frac{2k_F v_F + \omega}{k_x v_F}\right) - G\left(\frac{2k_F v_F - \omega}{k_x v_F}\right) \right] - \frac{2k_F}{\pi v_F \hbar}$, $e<0$ is the electron charge, $v_F$ is the Fermi velocity, $k_F$ is the Fermi wave vector, $G(z) = z\sqrt{z^2-1} - \ln(z+\sqrt{z^2-1})$ and $\omega < 2k_F v_F$. Taking into account the impact of collisions and introducing a phenomenological relaxation time $\tau$, the nonlocal conductivity should be modified as [50]

$$\sigma_g(\omega, k_x) = i\omega \frac{e^2}{k_x^2} \frac{(1+i/\omega\tau)\Pi(\omega+i/\tau, k_x)}{1+(i/\omega\tau)\Pi(\omega+i/\tau, k_x)/\Pi(0, k_x)}. \qquad (3)$$

Applying a DC bias voltage to the 1D graphene-based photonic crystal can generate drifting electrons in each graphene sheet (see the green arrow in Fig.1). In this scenario, the corresponding nonlocal conductivity can be modified as [37, 40]

$$\sigma_g^{drift}(\omega, k_x) = \frac{\omega}{\omega - k_x v_0} \sigma_g(\omega - k_x v_0, k_x) \qquad (4)$$

in which $v_0$ is the velocity of drifting electrons traveling along the $+x$ direction. This conductivity model can be obtained either by using a quantum mechanical method based on the self-consistent field approach [40], or by solving the semiclassical Boltzmann transport equation [41]. A detailed discussion about the variation of the current-carrying nonlocal graphene conductivity with either $k_x$ or $\omega$ can be found in Ref [39]. It should be noted that the Doppler-shifted factor $\omega - k_x v_0$ caused by the drifting electrons makes the nonlocal conductivity of graphene become nonreciprocal, i.e. $\sigma_g^{drift}(\omega, k_x) \neq \sigma_g^{drift}(\omega, -k_x)$. This means that time-reversal symmetry breaking in graphene-based plasmonic systems can be achieved by applying an electrical DC bias [37, 39].

## 3. NUMERICAL RESULTS AND DISCUSSIONS

The dispersion relation of TM-polarized propagating waves is given by solving Eq. (1). Here we consider $\varepsilon_d = 4$, $d = 100$ nm, fs, $\tau = 500$ m/s and $E_F = 0.1$ eV. These parameters are used throughout the rest of the paper. As shown in Fig. 2, the orange shaded regions and white areas represent allowed bands and forbidden bands, respectively. The green dots correspond to the band-crossing points in the inset of Fig. 2. Note that the upper part of these points is the photonic mode, and the lower part is the plasmonic mode originating from the coupling of SPPs between adjacent graphene layers [51-55]. The allowed band of the plasmonic mode shrinks progressively to the dispersion of a SPP for a single graphene layer (represented by blue curves in Fig. 2) as the period of the structure increases [55]. In Fig. 2(a), due to the absence of the applied electrical DC bias, i.e. $v_0 = 0$, the dispersion diagrams associated with the forward plasmonic mode $k_x > 0$ and backward plasmonic mode $k_x < 0$ are symmetric about the $k_x = 0$ axis. However, as illustrated in Fig.2(b), applying a DC bias voltage to the graphene-based photonic crystal and increasing the electron drift velocity to $0.85v_F$ can give rise to an extremely asymmetric band structure, namely, $\omega(k_x) \neq \omega(-k_x)$. Therefore, a strong nonreciprocal response of TM-polarized waves can exist in a graphene-based photonic crystal subjected to an applied DC bias.

In particular, expanding the dispersion equation Eq. (1) with respect to $k_z d$ and substituting $K_B = 0$ into Eq. (1), then we have [54]

$$k_z^2 = \frac{12}{d^2} \frac{d + 2\left(i\sigma_g^{drift} / 2\omega\varepsilon_0\varepsilon_d\right)}{d + 4d\left(i\sigma_g^{drift} / 2\omega\varepsilon_0\varepsilon_d\right)}. \tag{5}$$

Next, plugging $k_z = 0$, $k_x = -\sqrt{\varepsilon_d}\omega/c$ and $v_0 = 0.85v_F$ into Eq. (5), we thus obtain the operation frequency of the band-crossing point for the backward plasmonic mode $f \approx 8.96$ THz ($f = \omega/2\pi$). To further examine the feature of nonreciprocal responses, we plot the isofrequency contours (IFC) of the graphene-based photonic crystal with different electron drift velocity values in Fig. 3. For $v_0 = 0$ (see the green curves in Fig. (3), the hyperbolic contour of the plasmonic mode remains symmetric. Remarkably, the presence of the drifting electrons makes the hyperbolic contour become extremely asymmetric with respect to the $k_x = 0$ axis. As shown in Fig. 3(a), at the operation frequency $f = 8$ THz, the backward plasmonic mode ($k_x < 0$) undergoes the so-called topological transition with the increase of the drift velocity, and finally generates a closed ellipse [25]. In particular, such a closed ellipse will shrink to a point, i.e. the band-crossing point at the operation frequency $f = 8.96$ THz. As clearly seen in Fig.3(b), for the Bloch wave vector $K_B = 0$ and drift velocity $v_0 = 0$, the magnitudes of the wave vectors for the forward plasmonic mode and backward plasmonic mode are equivalent. However, as the drift velocity increases, in contrast to the forward plasmonic mode, the magnitude of the backward wave vector has a considerable increase. This means that the band gap, which is the magnitude of the difference between the wave vector of two plasmonic modes (forward mode and backward mode), becomes asymmetric about the $k_x = 0$ axis. Such an asymmetric band gap is useful for examining the one-way total transmission [17].

We now consider the effect of drifting electrons on the refraction angle. As schematically shown in the inset of Fig. 4(a) and 4(b), suppose a TM-polarized wave oblique incident from a dielectric material with permittivity $\varepsilon_i$ to the graphene-based photonic crystal with drifting electrons. The group velocity of the TM-polarized wave representing the direction of energy flow (see the red arrows in the Fig. 4) can be expressed as [28, 31]

$$V_g = \frac{\partial \omega}{\partial k_x}\hat{x} + \frac{\partial \omega}{\partial K_B}\hat{z} \quad (6)$$

As clearly shown in Fig. 4(b), when the electrons are traveling toward the $-x$ direction, the corresponding IFC is nearly flat. This means that incident waves with different wave vectors can be converted to TM-polarized waves with almost the same wave vectors. Such a flat IFC, also called as canalization, can be used to suppress diffraction [56,57]. According to the boundary continuity condition of the electromagnetic wave propagation, namely, $K_B = k_i\sqrt{\varepsilon_i}\sin\theta$ ($k_i = \omega/c$), the refraction angles of the wave vector and group velocity can be defined as [58]

$$\theta_k = \arctan\left(\frac{K_B}{k_x}\right), \quad (7)$$

and

$$\theta_g = \arctan\left(\frac{\partial\omega/\partial K_B}{\partial\omega/\partial k_x}\right) \quad (8)$$

respectively. Figure 4(c) shows the refraction angles of the wave vector and group velocity for different drift velocity values. when electrons move along the $+x$ direction, the drift velocity has an insignificant impact on the refraction angle. In contrast, for the electrons moving in the $-x$ direction, the refraction angle of the wave vector gradually decreases with the increase of the drift velocity. Notably, the refraction angle of the group velocity goes down from $-76°$ to nearly $0°$, a drop of $76°$. It should be noted that the refraction angle of the wave vector as well as group velocity drop to nearly $0°$, that is, drifting electrons driven by an DC electrical field can drag the wave vector and group velocity to the same direction, namely, the direction of the DC current ($+x$). As clearly seen in Fig. 4(d), the refraction angles of the wave vector and group velocity, represented by the red line and blue line respectively, almost overlap and maintain $0°$ as the incident angle increases. Thus, the refraction angles can be suppressed by the electrons moving with a large drift velocity when the normal component of the incident wave is antiparallel to the direction of the drifting electrons.

## 4. CONCLUSION

In summary, we have theoretically discussed the dispersion characteristics of a graphene-based photonic crystal that carries a DC current. Our numerical results indicate that drifting electrons driven by an external DC electric field can lead to asymmetric hyperbolic dispersion diagrams. Furthermore, when the normal component of the incident wave propagates along the direction of the DC current, the negative refraction will be suppressed. This means that the drifting electrons with a large enough drift velocity can drag the wave vector and group velocity to the same direction, which makes the electromagnetic wave energy flow along the direction of the DC current. Such properties not only can be used to design tunable nonreciprocal devices but also offer a novel route to control the light propagation without refraction.

**Acknowledgment**. We would like to acknowledge helpful discussions with T. A. Morgado and M. G. Silveirinha.

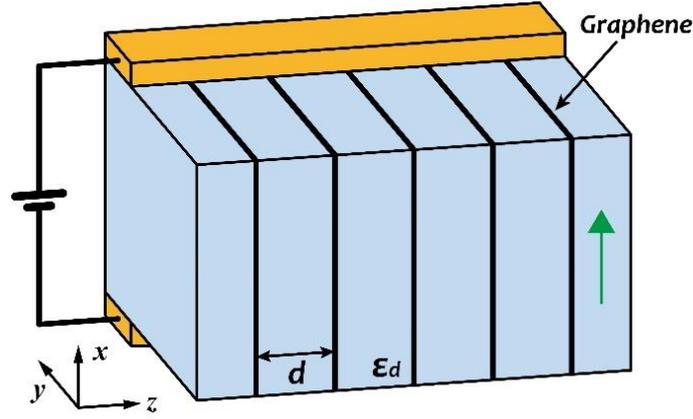

**Fig. 1.** Schematic of a graphene-based multilayer nanostructure with an applied electrical DC bias, where the green arrow indicates the direction of drifting electrons and each blue region represents a dielectric medium with relative permittivity $\varepsilon_d = 4$. The lattice period is $d = 100$ nm.

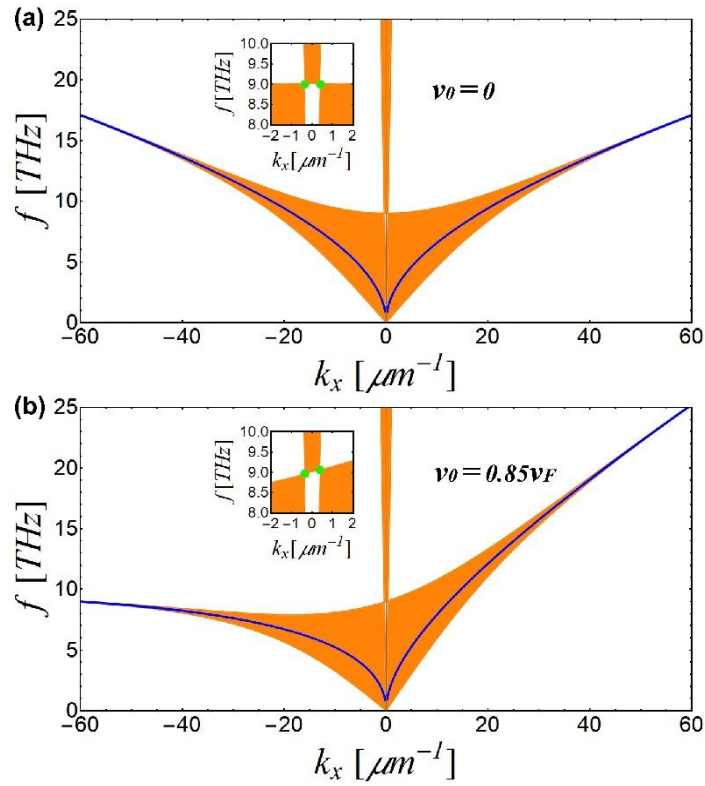

**Fig. 2.** Dispersion diagrams of TM-polarized propagating waves in a 1D graphene-based periodic nanostructure for (a) drift velocity $v_0 = 0$ and (b) drift velocity $v_0 = 0.85 v_F$. The band-crossing points marked by green dots are showed in the insets. The orange shaded regions correspond to the allowed bands. The blue line represents the dispersion of a single graphene sheet with the same dielectric permittivity as in Fig. 1. Other parameters are $E_F = 0.1\,\text{eV}$, $\tau = 500\,\text{fs}$, $d = 100$ nm and $v_F = 10^6$ m/s.

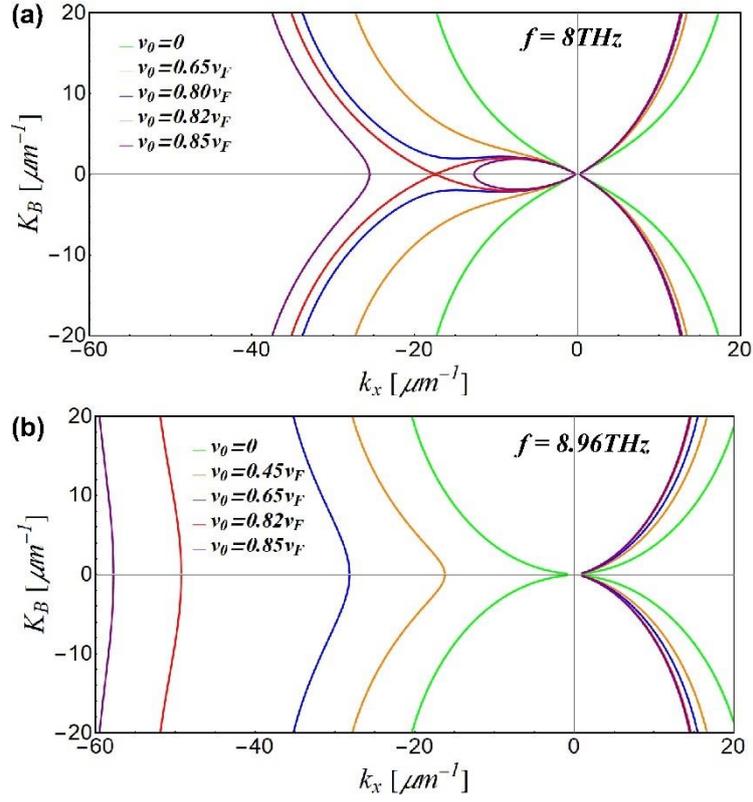

**Fig. 3.** Changes in the isofrequency contours of plasmonic modes caused by drifting electrons for (a) the operation frequency $f = 8$ THz and (b) the operation frequency corresponding to the band-crossing point $f = 8.96$ THz. Other parameters are the same as in Fig. 2.

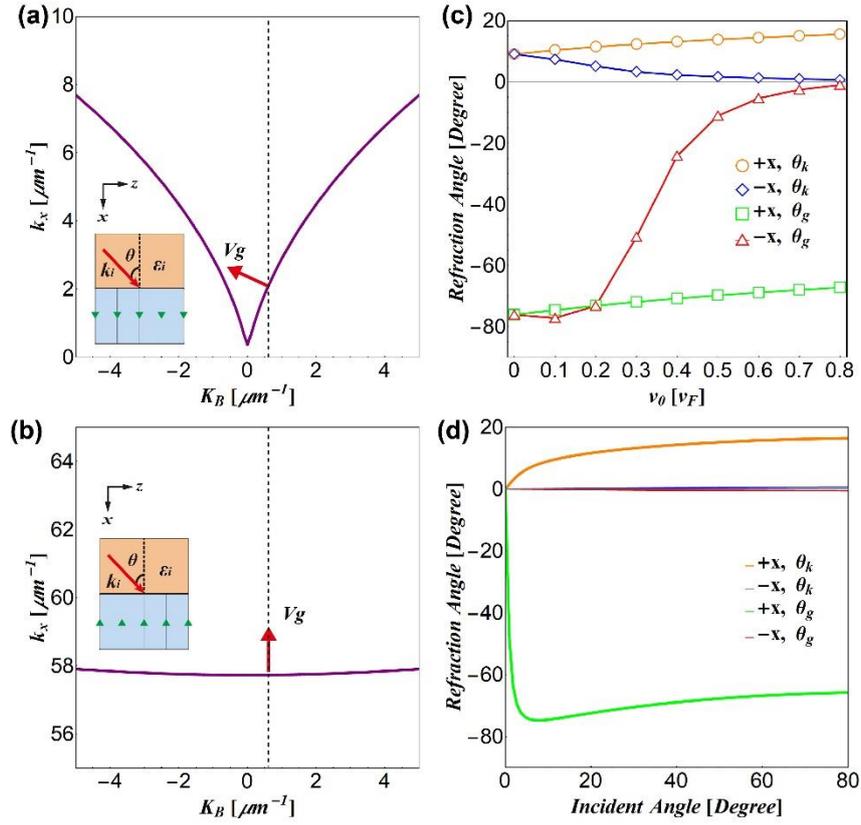

**Fig. 4.** (a) (b) Isofrequency contours of the graphene-based periodic nanostructure with drifting electrons traveling in different directions at a given drift velocity value $v_0 = 0.85 v_F$. Each inset shows a TM-polarized wave incident from a dielectric with permittivity $\varepsilon_i = 12$, where the green arrow represents the direction of the drifting electrons. The red arrow indicates to the direction of the group velocity, while the dashed line represents the continuity of $K_B = k_i \sqrt{\varepsilon_i} \sin\theta$ with $k_i = \omega/c$. (c) Refraction angles of the wave vector and group velocity vector for different drift velocity values at a given incident angle $\theta = 60°$. (d) Refraction angles for the wave vector and group velocity vector as functions of incident angles at a given drift velocity $v_0 = 0.85 v_F$. $+x$ ($-x$) corresponds to the direction of the drifting electrons. Operation frequency is 8.96 THz. Other parameters are the same as in Fig. 2.